# Electron supercollimation in graphene and Dirac fermion materials using one-dimensional disorder potentials

SangKook Choi[1], Cheol-Hwan Park[1,2], and Steven G. Louie[1*]

1. Department of Physics, University of California, Berkeley, and Materials Sciences Division, Lawrence Berkeley National Laboratory, Berkeley, CA 94720, USA

2. Department of Physics and Astronomy and Center for Theoretical Physics, Seoul National University, Seoul 151-747, Korea

Electron supercollimation, in which a wavepacket is guided to move undistorted along a selected direction, is a highly desirable property that has yet been realized experimentally. Disorder in general is expected to inhibit supercollimation. Here, we report a counter-intuitive phenomenon of electron supercollimation by disorder in graphene and related Dirac fermion materials. We show that one can use one-dimensional disorder potentials to control electron wavepacket transport. This is distinct from known systems where an electron wavepacket would be further spread by disorder and hindered in the potential fluctuating direction. The predicted phenomenon has significant implications in the understanding and applications of electron transport in Dirac fermion materials.

Dirac fermion materials are currently one of the most intensively investigated systems in condensed matter physics and materials science [1–6]. Its conical electronic structure near the Dirac points gives rise to massless neutrino-like two-dimensional (2D) electron states [7,8]. Due to the chiral nature of these Dirac fermion states, electrons interact with external potential in unusual ways, manifesting various interesting characteristics. In graphene, phenomena such as absence of backscattering by long-range potentials [9,10], Klein tunneling [11], weak-antilocalization [12–15], electron delocalization by one dimensional (1D) disorder [16–18], and supercollimation of electron beams by some specific 1D external periodic potentials [19] have been observed or predicted. Here we present another surprising, counter-intuitive electron transport phenomenon in graphene and related 2D Dirac fermion systems, made possible by the carriers' unique linear dispersion relation and chiral nature. We discovered that electron supercollimation can be induced by 1D disorder potentials. An electron wavepacket is guided to propagate virtually undistorted along the fluctuating direction of the external 1D disorder potential, independent of its initial motion, as long as the disorder is large enough to produce a wedge-like dispersion in the bandstructure within which the $k$-components of the wavepacket are contained. To our knowledge, this phenomenon was not known in any medium previously. Further, we find, for graphene in an external periodic potential that doesn't satisfy the supercollimation condition predicted in Ref. [19], addition of disorder would enhance collimation. The more is the disorder, the better is the supercollimation. This robust novel phenomenon has significant implications in the fundamental understanding of transport in graphene, as well as in other materials with Dirac cone physics (such as surface states of topological insulators [5] or possibly certain photonics crystals [6]), and has the potential to be exploited in the design of devices based on these materials.

We first discuss the predicted supercollimation by disorder in Dirac fermion materials using results from direct simulations (Fig. 1) and then derived the phenomenon from perturbation theory. For the low energy carriers in graphene (Fig. 2(a)) and an external potential $V(x)$ that depends only on $x$, we may set up an effective Hamiltonian for the electronic states [8]

$$H = v_0 \sigma_x p_x + v_0 \sigma_y p_y + V(x) I . \tag{1}$$

Here $v_0$ is the band velocity of the electron, $\sigma_i$ is the i-component Pauli matrix, $p_i$ is the i-direction momentum operator, and $I$ is the identity matrix in the space describing the pseudo-spin components of the electron wavefunction. We carried out direct numerical simulations on the Hamiltonian in Eq. (1) using a spatially-correlated Gaussian disorder potential for $V(x)$. Such a disorder potential is characterized by a two-point correlation function having $\overline{V(x_1)V(x_2)} = \Delta^2 e^{-|x_1-x_2|/l_c}$ where $\Delta$ is the magnitude of the disorder fluctuation and $l_c$ is the disorder correlation length. The overline represents ensemble-averaged value. Fig. 1(a) shows one realization of $V(x)$ in unit of $\Delta$. The potential is spatially correlated so that on average it is nearly the same value within a length scale of $l_c$.

Figure 1(b)-(d) demonstrate supercollimation in a Gaussian wavepacket propagation simulation. Using 60 different realizations of the disorder potential $V(x)$ with $l_c \Delta = 4\pi \hbar v_0$, we numerically calculated the electron density $\overline{\rho(r,t)}$ using an initial Gaussian density packet with initial center of mass wavevector $k_0 = \pi/5l_c$ and a half width of $r_0 = 5l_c$. Figure 1(c) and 1(d) show the evolution of the electron density $\overline{\rho(r,t)}$ from the initial electron density shown in Fig. 1(b). In the absence of a disorder potential, the Gaussian wavepacket propagates along the initial center of mass wavevector direction marked by the white arrow and spreads sideway. Its spread angle at which the electron density is half the maximum is $48.6°$. With the 1D disorder potential

$V(x)$, the electron package propagates nearly un-spread along the potential fluctuation direction, which is $x$, regardless of the initial velocity direction. The spread angles are $0.5°$ and $0.7°$ at an incident angle (measured from the x-axis) of $0°$ and $45°$, respectively. A very tiny fraction of the electron density forms a supercollimated trail (barely visible in Fig. 1(d)), which increases with increasing incident angle. The same behavior is observed in simulations within a tight-binding Hamiltonian framework for graphene (see Supplemental Material [20]).

We present now an analytic derivation of the phenomenon. We separate the Hamiltonian in Eq. (1) into two terms, $H = H_0 + H_1$, with $H_0 = v_0 \sigma_x p_x + V(x)I$ and $H_1 = v_0 \sigma_y p_y$. If $H_0$ dominates over $H_1$ (to be defined more precisely below) as in the case of an extended low-energy wavepacket in real space in a disorder $V(x)$, we may regard $H_1 = v_0 \sigma_y p_y$ as a perturbation.

We first show that the electron dynamics in 2D governed by $H_0 = v_0 \sigma_x p_x + V(x)I$ alone yields supercollimation along the $x$ direction. The term $v_0 \sigma_x p_x$ has two eigenstates ($s = \pm 1$) as shown in Fig. 2(b). By a unitary transformation of $U = \frac{1}{\sqrt{2}}\begin{pmatrix} 1 & 1 \\ 1 & -1 \end{pmatrix}$, $U^\dagger v_0 p_x \sigma_x U$ is diagonal with eigenvalues $s\hbar v_0 k_x$ with $s = \pm 1$, and eigenvectors $\frac{1}{\sqrt{A}}\begin{pmatrix} e^{i\mathbf{k}\cdot\mathbf{r}} \\ 0 \end{pmatrix}'$ and $\frac{1}{\sqrt{A}}\begin{pmatrix} 0 \\ e^{i\mathbf{k}\cdot\mathbf{r}} \end{pmatrix}'$, respectively, where $A$ is the area of the sample and the prime notation here indicates a matrix or vector in the unitary-transformed or pseudo-spin basis. These are chiral states moving forward ($s=1$) or backward ($s=-1$) with a speed of $v_0$ and a pseudo spin aligned along the propagation direction. The retarded Green's function $G_0'$ of $H_0' = U^\dagger H_0 U$ in coordinate space is given by

$$G'_0(\mathbf{r},\mathbf{r}',t) = \frac{1}{i\hbar}\theta(t)\delta(y-y')\begin{pmatrix} \delta(x-x'-v_0 t)\alpha(x,x') & 0 \\ 0 & \delta(x'-x-v_0 t)\alpha(x',x) \end{pmatrix} \quad (2)$$

with

$$\alpha(x,x') = \exp\left(\frac{1}{i\hbar v_0}\int_{x'}^{x} V(x_1)dx_1\right) \quad (3)$$

(see Supplemental Material [20] for the derivation of Eq. (2); $G'_0$ is consistent with the transfer matrix in Ref. [21].) The Green's function determines the time evolution of the electron wavefunction and density through

$$\psi'(\mathbf{r},t) = \int d\mathbf{r}' i\hbar G'(\mathbf{r},\mathbf{r}',t)\psi'_0(\mathbf{r}',t=0), \quad (4)$$

and

$$\rho(\mathbf{r},t) = tr[\psi'(\mathbf{r},t)\psi'^{\dagger}(\mathbf{r},t)], \quad (5)$$

where the trace is defined with respect to the $2\times 2$ pseudo-spin subspace. (We recall that $\psi'(\mathbf{r},t)$ is a 2-component spinor function and the total density $\rho(\mathbf{r},t)$ is a sum over densities from the two components.) As seen from the diagonal-matrix form of $G'_0(\mathbf{r},\mathbf{r}',t)$ in Eq. (2), scattering between two states with different chirality (or group velocity) is not allowed for any arbitrary external potential $V(x)$, if we neglect $H_1$. Consequently, for the Hamiltonian $H'_0$, the amplitude of any initial wavefunction $\psi'(\mathbf{r},t=0)$ with pseudo-spin $s$ moves at a velocity of $sv_0$, maintaining its initial shape, although the phase of the wave function is changed by the interaction with the potential $V(x)I$. The electron density of a wavepacket with a pseudo-spin $s$ thus also propagates with a velocity of $sv_0$ along the $x$ direction, maintaining its original shape at $t=0$, again, if $H'_1$ is neglected. To illustrate this point, if we take an initial Gaussian wave packet with initial center of mass wavevector $\mathbf{k}_0$ and a half width of $\sqrt{2}r_0$,

$$\psi'(\mathbf{r},t=0) = \frac{1}{\sqrt{2\pi}r_0}\begin{pmatrix}1\\0\end{pmatrix}'\exp\left(-\frac{r^2}{4r_0^2}+i\mathbf{k}_0\cdot\mathbf{r}\right). \tag{6}$$

then, as a function of time, the electron density is given by (from Eqs. (4) and (5))

$$\rho^{(0)}(\mathbf{r},t) = \frac{1}{2\pi r_0^2}\exp\left(-\frac{|\mathbf{r}-\mathrm{v}_0t\hat{\mathbf{x}}|^2}{2r_0^2}\right). \tag{7}$$

The disorder potential $V(x)$ generates a random phase accumulation for the electron which may be loosely thought of as an effective elastic mean free path $l_s$ or elastic collision time $\tau$ for electrons governed by $H_0'$. The quantity $l_s$ may be extracted from $\overline{G_0'}$. In the expression for $G_0'$ given by Eq. (2), the quantity $\alpha(x,x')$ incorporate all the effects of $V(x)$. For a random potential, translation symmetry is restored by ensemble average [22], so that $\overline{\alpha(x,x')} = \overline{\alpha(x-x')}$. The form of $\alpha$ in Eq. (3) dictates that $\overline{\alpha(x-x')}$ has its maximum at $x=x'$, and decreases as $|x-x'|$ increases since the phase of $\alpha(x,x')$ fluctuates from one member to another in an ensemble. If we assume that $\overline{\alpha(x,x')}$ decays with a full-width-half-maximum of $l_s$, then $\overline{G_0'(\mathbf{r}-\mathbf{r}',t)}$ decays with the same mean-distance $l_s$. The effective elastic collision time $\tau$ is obtained by considering $\overline{G_0'}$ in Fourier space. A Fourier transform of Eq. (2) yields

$$\overline{G_0'(\mathbf{k},\omega)} = \int dE'\frac{1}{\hbar\omega-E'+i\eta}\begin{pmatrix}A_0(E'-\hbar\mathrm{v}_0k_x) & 0\\0 & A_0(E'+\hbar\mathrm{v}_0k_x)\end{pmatrix} \tag{8}$$

with

$$A_0(E) = \frac{1}{2\pi\hbar\mathrm{v}_0}\int dx\,\overline{\alpha(x)}\exp\left(i\frac{E}{\hbar\mathrm{v}_0}x\right). \tag{9}$$

The function $A_0(E'-s\hbar\mathrm{v}_0k_x)$ here plays the role of the spectral function $A_0(s,\mathbf{k},\omega)$. Due to the decay of $\overline{\alpha(x)}$, $A_0(s,\mathbf{k},\omega)$ is maximum at $\omega = s\mathrm{v}_0k_x$ and has a finite width, owing to the finite effective elastic collision time, which is independent of the momentum $\boldsymbol{k}$. From the full-width-half-maximum of $A_0(s,\boldsymbol{k},\omega)$, we can deduce $\tau$. For example, for a spatially correlated Gaussian disorder, one obtains:

$$\overline{\alpha(x)} = \exp\left[-\left(\frac{l_c \Delta}{\hbar v_0}\right)^2 \left\{\exp\left(-\frac{|x|}{l_c}\right) - 1 + \frac{|x|}{l_c}\right\}\right]. \tag{10}$$

Let us now consider the effects of $H_1 = v_0 \sigma_y p_y$ and show that electron supercollimation still persists over a large distance $L_0$. We show this by examining the time evolution of the electron density $\rho(\boldsymbol{r},t)$ by the full Hamiltonian $H' = U^\dagger H U$ from a series expansion of the wavefunction up through third order in $H_1' = U^\dagger H_1 U = -v_0 p_y \sigma_y$, i.e., $\psi'(\boldsymbol{r},t) \approx \sum_{i=0}^{3} \psi'^{(i)}(\boldsymbol{r},t)$ where the change in the wavefunction in ith order is given by $\psi'^{(i)}(\boldsymbol{r},t)$. Then,

$\rho(\boldsymbol{r},t) \approx \sum_{i=0}^{3} \rho^{(i)}(\boldsymbol{r},t)$ with $\rho^{(i)}(\boldsymbol{r},t) \equiv \sum_{\substack{j,k=0 \\ j+k=i}}^{3} \rho_{jk}^{(i)}(\boldsymbol{r},t)$ and $\rho_{jk}^{(j+k)}(\boldsymbol{r},t) = tr[\psi'^{(j)}(\boldsymbol{r},t)\psi'^{(k)\dagger}(\boldsymbol{r},t)]$. The zeroth order term $\rho^{(0)}$, given by Eq. (7), is already shown to be a collimated electron density with unchanging shape. Up to the third order in $H_1'$, it is straight-forward to show that all terms in the above expression for $\rho(\boldsymbol{r},t)$, except one, retain the initial extent of the wavepacket and move along the $x$ direction with the same velocity. Only the $\rho_{11}^{(2)}(\boldsymbol{r},t)$ term shows shape deviation, but it still does not spread along the $y$ direction and is collimated to move along the $x$ direction (for the details, see Supplemental Material [20]). To illustrate this, for $r_0 > l_s$ and with an initial wavepacket given by Eq. (6), $\overline{\rho}_{11}^{(2)}(\boldsymbol{r},t)$ is

$$\overline{\rho}_{11}^{(2)}(\boldsymbol{r},t) \approx \frac{l_s}{2\sqrt{2\pi} r_0}\left(k_{0y}^2 + \frac{y^2}{4r_0^4}\right) e^{-y^2/2r_0^2}\left(-Erf\left(\frac{x - v_0 t}{\sqrt{2} r_0}\right) + Erf\left(\frac{x + v_0 t}{\sqrt{2} r_0}\right)\right). \tag{11}$$

This corresponds to a strip of density of width $2r_0$ determined by the initial wavepacket but extended from $v_0 t$ to $-v_0 t$ in the $\hat{x}$ direction. If we compare $\int d\mathbf{r} \overline{\rho_{11}^{(2)}(\mathbf{r},t)}$ with $\int d\mathbf{r} \rho^{(0)}(\mathbf{r},t)$, $\int d\mathbf{r} \overline{\rho_{11}^{(2)}(\mathbf{r},t)} < \int d\mathbf{r} \rho^{(0)}(\mathbf{r},t)$ for $(2k_{0y}^2 + 1/(2r_0^2))l_s v_0 t < 1$, giving rise to supercollimation with little diminishment of the intensity of the original Gaussian profile over a distance of roughly $L_0 = v_0 t = 1/\left(2l_s k_{0y}^2 + l_s/(2r_0^2)\right)$. For example, for a disorder potential that gives a broadening of 0.2 eV in the spectral function, a wavepacket with $r_0 > 40 nm \approx 250 a_{cc}$ and a center of mass wavevector such that $\hbar v_0 k_{0y} < 0.01 eV$ will undergo supercollimation for nearly a micrometer.

We propose a possible experiment to demonstrate the predicted electron supercollimation phenomenon by measuring the conductance $G$ in a geometry shown in Fig. 3(a). In this set up, graphene or a Dirac fermion material is in contact with two electrodes that are separated at a distance $L$ along the $\hat{\mu}$ direction. This direction is at an angle $\theta$ with respect to the 1D potential fluctuation direction $\hat{x}$. The conductance $G$ between the two electrodes is, according to the Kubo formula [23,24],

$$G(L,\theta) = \int d\mathbf{r} d\mathbf{r}' \sigma_{\mu\mu}(\mathbf{r},\mathbf{r}',E_F)\delta(\mu)\delta(\mu'-L), \qquad (12)$$

with conductivity

$$\sigma_{\mu\mu}(\mathbf{r},\mathbf{r}',E_F) = \frac{\pi\hbar}{(2\pi i)^2} tr\left[ j_\mu(\mathbf{r})(G^{R-A}(\mathbf{r},\mathbf{r}',E_F)) j_\mu(\mathbf{r}')(G^{R-A}(\mathbf{r}',\mathbf{r},E_F)) \right], \qquad (13)$$

where $\mathbf{j} = e v_0 \boldsymbol{\sigma}$ [25]. The quantity $G^{R-A}$ is defined as $G^{R-A} = G^R - G^A$ with $G^R$ and $G^A$ being the retarded and advanced Green's functions, respectively. If we expand the ensemble-average conductance $\overline{G(L,\theta)}$ up to and including the first order term in $H_1$ (see Supplemental Material [20]),

$$\overline{G(L,\theta)} \propto \cos^2\theta + \sin^2\theta \exp[-2L/(l_s \cos\theta)]\cos(2E_F L/(\hbar v_0 \cos\theta)). \qquad (14)$$

This is dramatically distinct from that of a gated pristine graphene in the ballistic regime, in which case the conductance $G(L,\theta)$ is constant regardless of the orientation angle $\theta$.

The ensemble-average dispersion relation of the electrons in graphene is strongly and anisotropically renormalized in the presence of the random potential $V(x)$ for $|k_y|<1/l_s$ (for details, see Supplemental Material [20]), forming a wedge-like structure for the energy surface $E(\mathbf{k})$. We demonstrate this effect by calculating the ensemble-average spectral function. Using 60 different realizations of the disorder with $l_c\Delta = 4\pi\hbar v_0$, we numerically calculated $\text{Im}\,G'(\mathbf{k},\omega)$. Figure 4 shows the 60-ensemble-average spectral function, $A(\mathbf{k},E) = -tr\,\text{Im}\,\overline{G'(\mathbf{k},E)}/\pi$ where the trace is with respect to the $2\times 2$ pseudo-spin subspace. Along the $\mathbf{k}=(k_x,0)$ line, shown in Fig. 4(a), the dispersion relation is linear and it follows the $E=\pm\hbar v_0 k_x$ lines, which is the dispersion relation of the pristine system. However, along the $\mathbf{k}=(0,k_y)$ line, shown in Fig. 4(b), the bandstructure is strongly renormalized near the Dirac point and becomes flat. The anisotropic renormalization of the band structure can be demonstrated more clearly by a contour plot of $A(\mathbf{k},E)$ on the $k_x$-$k_y$ plane with $E=2\hbar v_0/l_s$, as shown in Fig. 4(c). On this constant energy plane, constant amplitude lines of $A(\mathbf{k},E)$ are oval-shaped and stretched along the $k_y$ direction. For spatially-correlated Gaussian disorder potentials, we can evaluate explicitly the spectral function of $\overline{G}'_0$ from the Fourier transform of Eq. (10). For this particular kind of disorder, the lineshape of the spectral function at different $k_x$ with $|k_{0y}|<1/l_s$ is identical and $\tau \sim \hbar/\Delta$ (if $l_c > \hbar v_0/\Delta$). As shown in Fig. 4(d), the line shape from Eqs. (9) and (10) matches well with numerically simulated line shape from $\overline{G}'$ at various $k_x$ with $k_y = 0$. However, as $k_y$ increases (at $k_x = 0$), the numerically calculated spectral function deviates from the lineshape from $\overline{\alpha(x)}$ owing to the effect of $H_1'$.

Electron beam supercollimation has been predicted theoretically in certain special graphene superlattices (SGS): a graphene sheet modulated by an 1D periodic potential satisfying certain specific conditions [19]. In the experimental realization of such SGS (e.g., using substrate [26], controlled adatom deposition [27], ripples [28] under perpendicular electric field [29,30], or gating [31]), it would be unavoidable to have some disorders in the external potential, which previously thought might impede the supercollimation effect. However, we found in this study that 1D disorder along the periodic potential modulation direction in fact enhances supercollimation for an external periodic potential, with such enhancement occurring even if the external potential does not exactly satisfy the SGS supercollimation condition. For a detailed discussion and numerical simulation of this phenomenon, please see Supplemental Material [20].

In summary, through perturbation theory analysis and numerical simulations, we have discovered a highly counter-intuitive phenomenon of electron supercollimation via 1D disorder potential in graphene and other Dirac fermion materials. To our knowledge, this phenomenon is not seen in any other systems.

*e-mail: sglouie@berkeley.edu

**Acknowledgments**

We wish to thank M. L. Cohen and G. Y. Cho for discussions. Analytical calculations was supported by National Science Foundation Grant No. DMR10-1006184. Numerical calculation was supported by the Theory Program at the Lawrence Berkeley National Lab funded by the Director, Office of Science, Office of Basic Energy Sciences, Materials Sciences and Engineering Division, U.S. Department of Energy under Contract No. DE-AC02-05CH11231. Computational resources have been provided by the DOE at Lawrence Berkeley National Laboratory's NERSC facility. S.G.L. acknowledges support by a Simons Foundation Fellowship in Theoretical Physics and C.-H.P. by Korean NRF funded by MSIP (Grant No. NRF-2013R1A1A1076141).

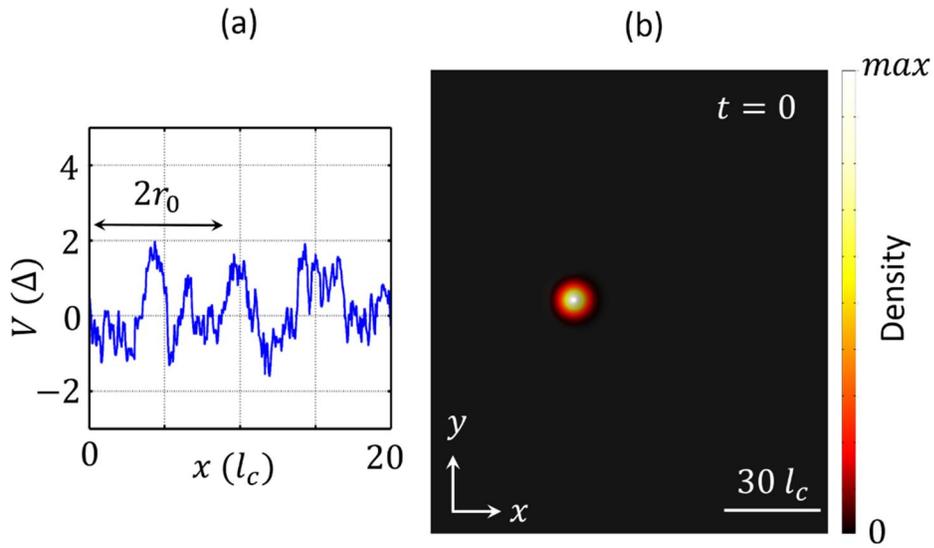

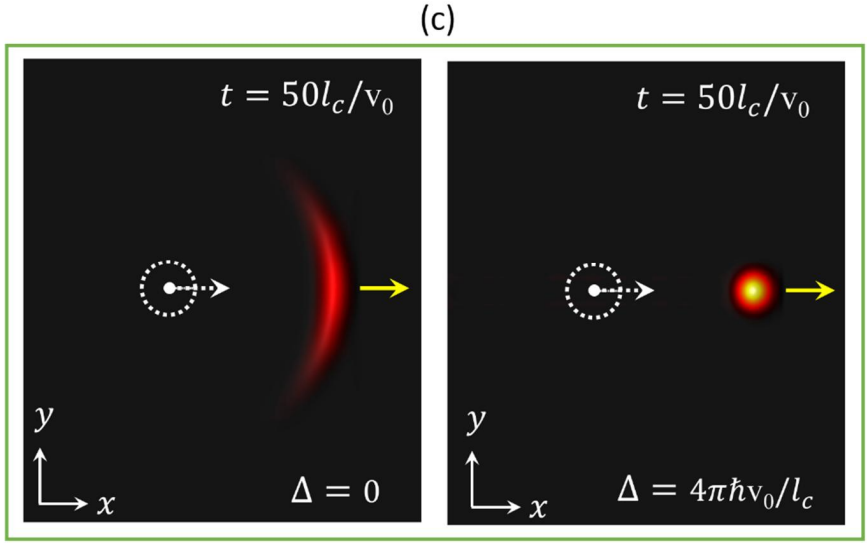

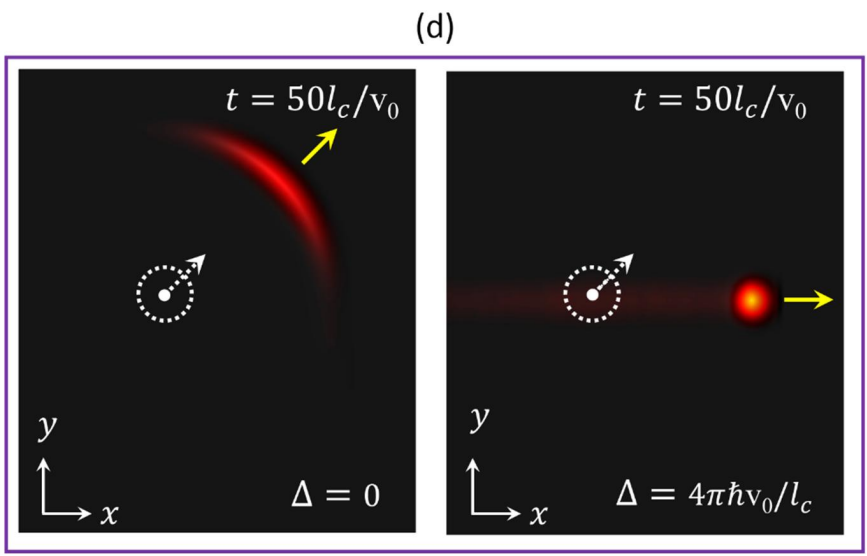

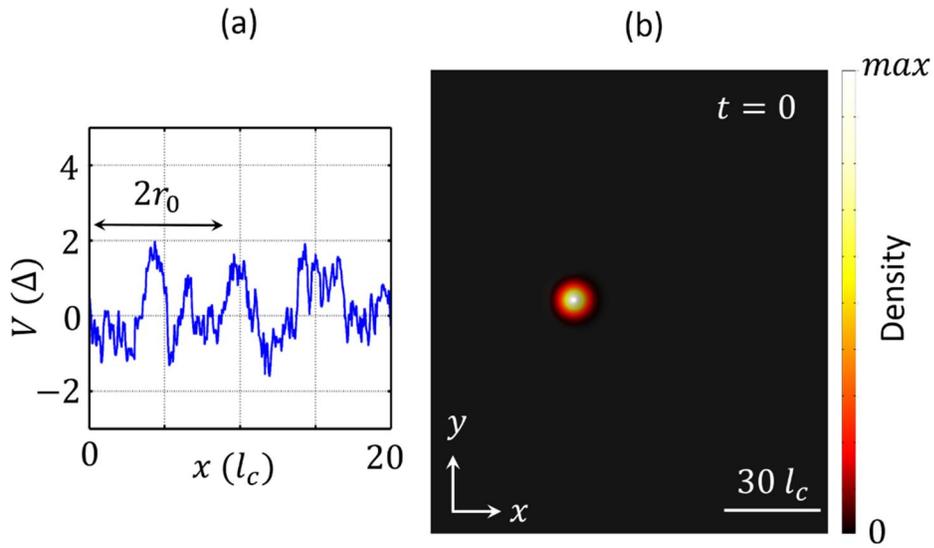

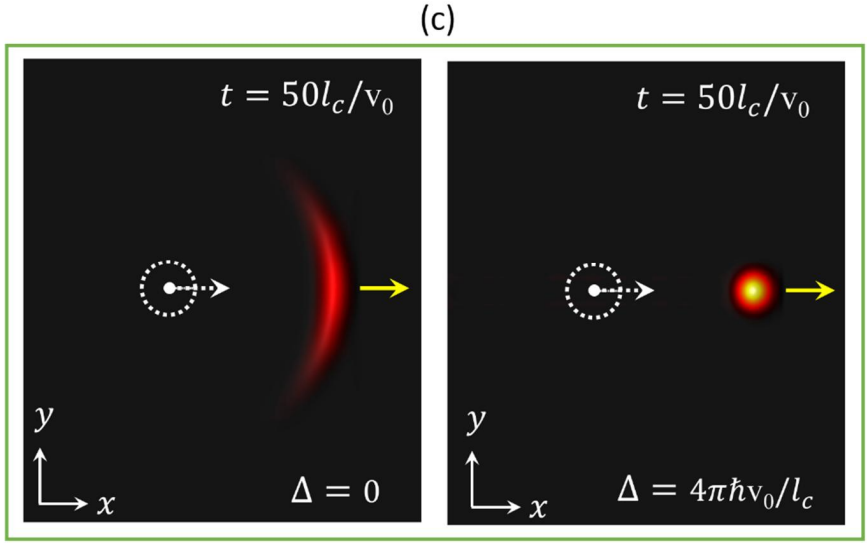

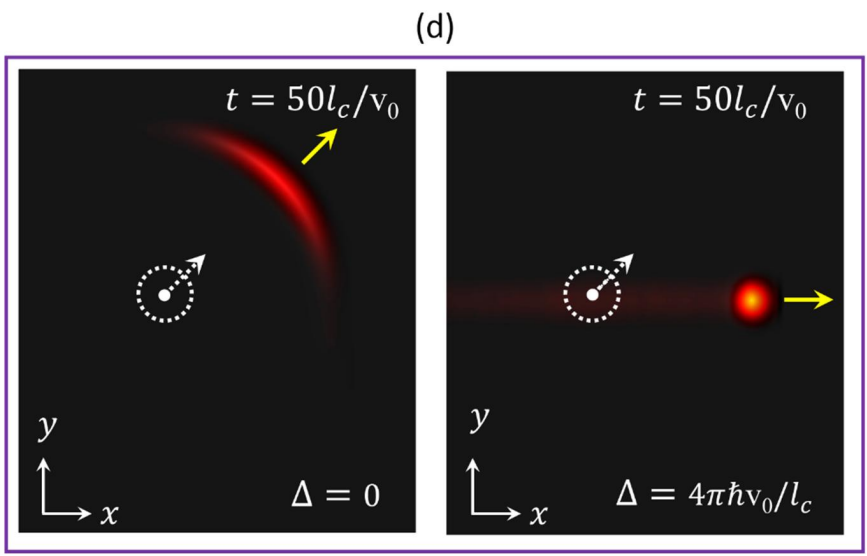

Fig. 1 (color online). (a) A realization of spatially-correlated Gaussian disorder potential $V(x)$ with a magnitude $\Delta$ and correlation length $l_c$. (b) Initial wavepacket with electron density in a Gaussian shape in coordinate space with initial center of mass wavevector $k_0 = \pi/5l_c$ and a half width of $r_0 = 5l_c$. (c)-(d) Electron density distribution in coordinate space at time $t = 50l_c/v_0$ in pristine system (left panel) and in disordered system (right panel) with initial center of mass wavevector direction (white arrow) pointing with respect to the x axis at $0°$ (c) and $45°$ (d).

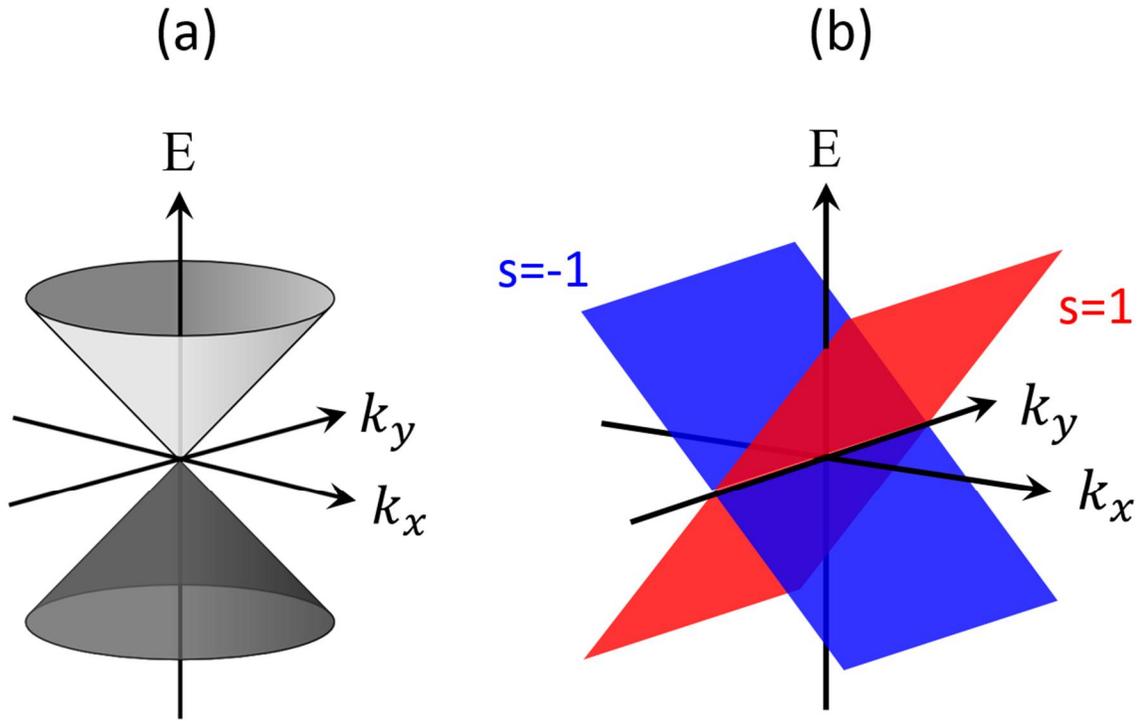

Fig. 2 (color online). (a) Low-energy electronic bandstructure of graphene near a Dirac point. (b) Electronic bandstructure of an initial 2D model Hamiltonian, $H_{in} = v_0 \sigma_x p_x$, where $v_0$ is the band velocity, $\sigma_x$ is the $x$-component Pauli matrix and $p_x$ is the x-direction momentum operator. This model Hamiltonian generates two chiral eigenstates which correspond to forward-moving ($s=1$) and backward-moving ($s=-1$) states with a speed of $v_0$ and a pseudo-spin parallel to $s\hat{x}$.

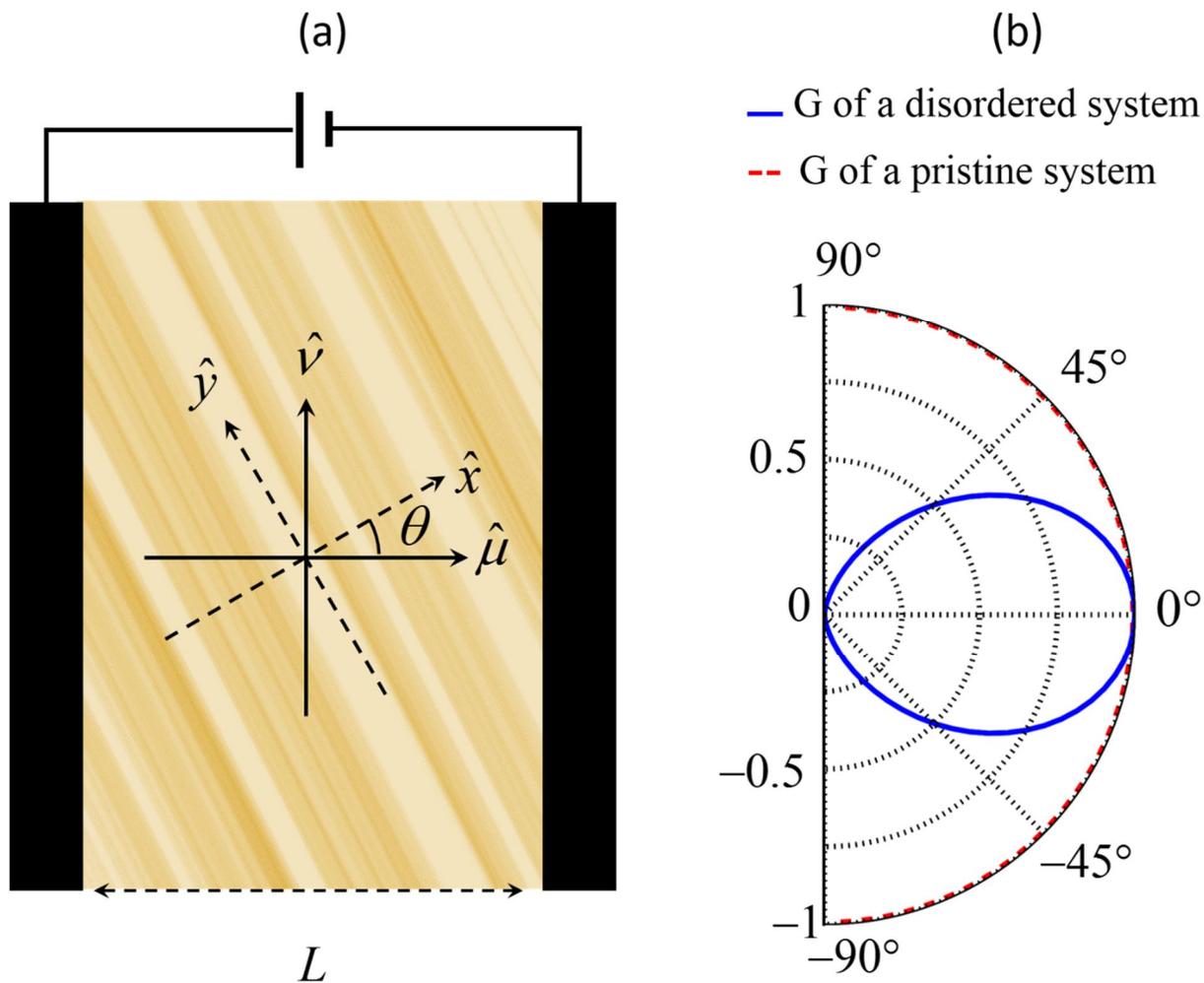

Fig. 3 (color online). (a) Schematic diagram of experimental setup for proposed conductance measurement. Two electrodes are in contact with graphene under 1D disorder potential fluctuating along the $\hat{x}$ direction. The electrodes are separated by a distance $L$ along the $\hat{\mu}$ direction. (b) Calculated conductance $G(L,\theta)$ (in a unit of $2N_v e^2/h$ where $N_v$ is the number of subbands due to the confinement along the $v$ direction at energy $E_F$) as a function of the angle $\theta$ in a pristine system in the ballistic regime (red line) and in a system with 1D disorder potential ($l_s \ll L$) shown in (a) (blue line).

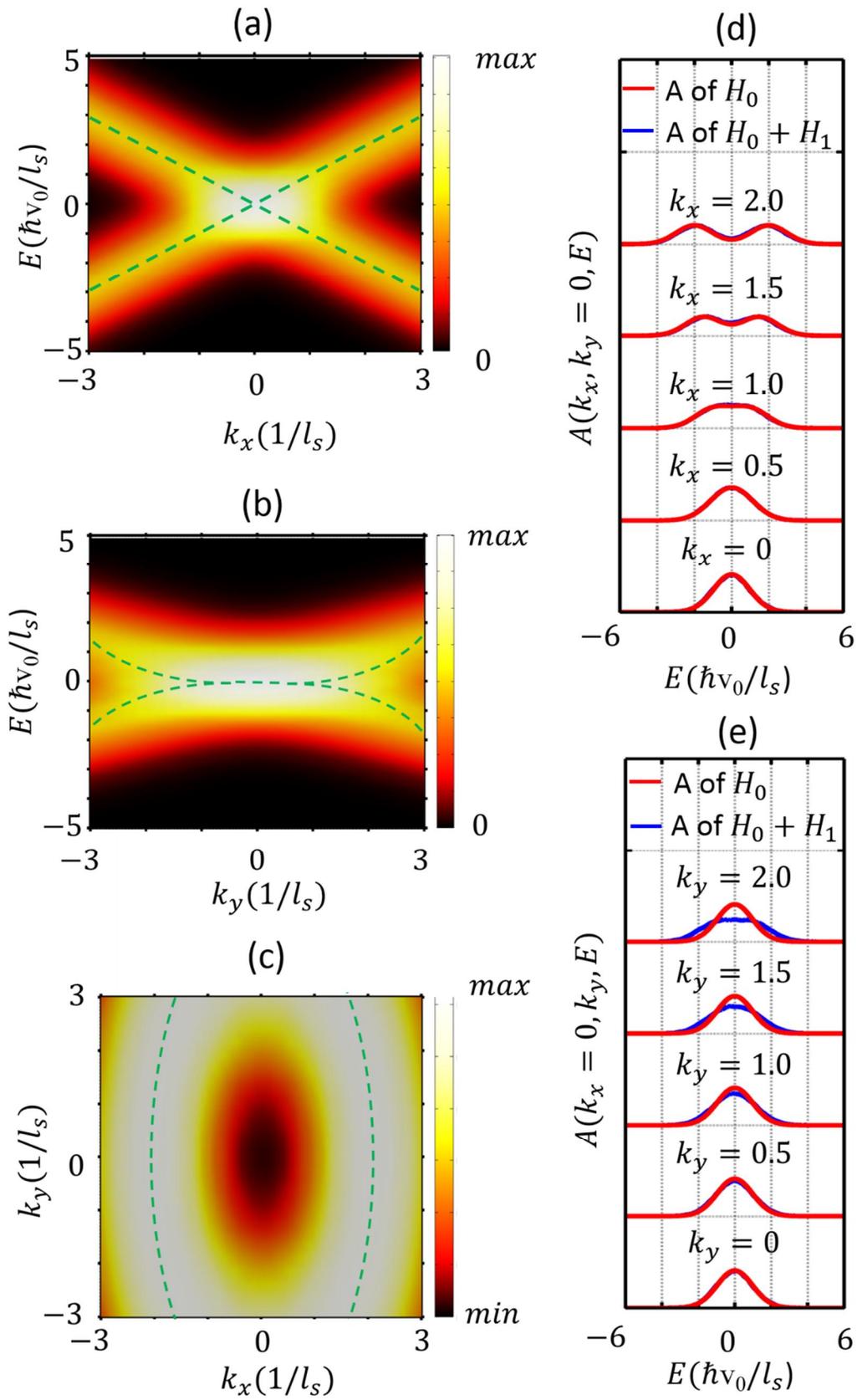

Fig. 4 (color online) (a)-(c) Numerically evaluated spectral function, $A(\mathbf{k},E) = -tr\,\mathrm{Im}\,\overline{G'(\mathbf{k},E)}/\pi$, for $l_c\Delta = 4\pi\hbar v_0$ along the $k_y = 0$ line (a), along the $k_x = 0$ line (b), and on the $E = 2\hbar v_0 / l_s$ plane with $l_s = \dfrac{\hbar v_0}{\Delta}$ (c). (d)-(e) The lineshapes of the spectral function of $H_0 = v_0 \sigma_x p_x + V(x)I$ (red lines) and of $H_0 + H_1$ (blue lines) with $H_1 = v_0 \sigma_y p_y$ at various $k_x$ with $k_y = 0$ (d), and at various $k_y$ with $k_x = 0$ (e).

# SUPPLEMENTAL MATERIAL

# Electron Supercollimation in Graphene and Dirac Fermion Materials Using One-dimensional Disorder Potentials

SangKook Choi[1], Cheol-Hwan Park[1,2], and Steven G. Louie[1]*

1. Department of Physics, University of California, Berkeley and Materials Sciences Division, Lawrence Berkeley National Laboratory, Berkeley, CA 94720, USA

2. Department of Physics and Astronomy and Center for Theoretical Physics, Seoul National University, Seoul 151-747, Korea

*e-mail: sglouie@berkeley.edu

1. Retarded Green's function $G'_0(r,r',t)$ of $H'_0$

We derived the exact retarded Green's function $G'_0(r,r',t)$ of $H'_0$ with $H'_0 = U^\dagger H_0 U$ and $H_0 = v_0 \sigma_x p_x + V(x) I$ using a perturbation expansion (see main text). First, the retarded Green's function of the first or kinetic term in $H'_0$ is

$$G'_{0T}(r,r',t) = \frac{1}{i\hbar}\delta(y-y')\theta(t)\begin{pmatrix} \delta(x-x'-v_0 t) & 0 \\ 0 & \delta(x'-x-v_0 t) \end{pmatrix}. \tag{s1}$$

The retarded Green's function of $H'_0$ to infinite order in $V(x)$ is,

$$G'_0(r,r',t)$$
$$= G'_{0T}(r,r',t)$$
$$+ \sum_{n=1}^{\infty} \int dr_1 \cdots dr_n dt_1 \cdots dt_n G'_{0T}(r,r_1,t-t_1) V(r_1) I\, G'_{0T}(r_1,r_2,t_1-t_2) \cdots V(r_n) I\, G'_{0T}(r_n,r',t_n)$$

$$= G'_{0T}(r,r',t) + \sum_{n=1}^{\infty} \frac{1}{(i\hbar)^{n+1}} \int_0^t dt_1 \int_0^{t_1} dt_2 \cdots \int_0^{t_{n-1}} dt_n \delta(y-y')\theta(t)$$

$$\times \begin{pmatrix} V(x-v_0(t-t_1)) \cdots V(x-v_0(t-t_n)) & 0 \\ \times \delta(x-x'-v_0 t) & \\ & V(x+v_0(t-t_1)) \cdots V(x+v_0(t-t_n)) \\ 0 & \times \delta(x'-x-v_0 t) \end{pmatrix}. \tag{s2}$$

Since $V(x)$ at different positions commute with each other [1],

$$G'_0(r,r',t)$$
$$= G'_{0T}(r,r',t) + \sum_{n=1}^{\infty} \frac{1}{(i\hbar)^{n+1} n!} \int_0^t dt_1 \int_0^t dt_2 \cdots \int_0^t dt_n \delta(y-y')\theta(t)$$

$$\times \begin{pmatrix} V(x-v_0(t-t_1)) \cdots V(x-v_0(t-t_n)) & 0 \\ \times \delta(x-x'-v_0 t) & \\ & V(x+v_0(t-t_1)) \cdots V(x+v_0(t-t_n)) \\ 0 & \times \delta(x'-x-v_0 t) \end{pmatrix} \tag{s3}$$

$$= \frac{1}{i\hbar}\theta(t)\delta(y-y')\begin{pmatrix} \delta(x-x'-v_0 t) e^{\frac{1}{i\hbar v_0}\int_{x'}^{x} V(x_1) dx_1} & 0 \\ 0 & \delta(x'-x-v_0 t) e^{\frac{1}{i\hbar v_0}\int_{x}^{x'} V(x_1) dx_1} \end{pmatrix},$$

which satisfies $i\hbar \partial_t G_0'(r_1, r', t) - H_0'(r) G_0'(r, r', t) = \delta(r - r') \delta(t) I$

2. Series expansion of $\rho(r,t)$ in $H_1'$ ($H_1' = U^\dagger H_1 U = -v_0 p_y \sigma_y$)

Suppose that we expand $\rho(r,t)$ and $\psi'(r,t)$ to third order in $H_1'$:

$$\psi'(r,t) \approx \sum_{i=0}^{3} \psi'^{(i)}(r,t) \tag{s4}$$

and

$$\rho(r,t) \approx \sum_{i=0}^{3} \rho^{(i)}(r,t), \tag{s5}$$

where $\psi'^{(i)}(r,t)$ and $\rho^{(i)}$ are the ith order terms in $H_1'$. Substituting these expansions to Eq. (5) in the main text and arranging them according to powers in $H_1'$,

$$\rho^{(0)}(r,t) = tr\left[\psi^{(0)}(r,t)\psi^{(0)\dagger}(r,t)\right], \tag{s6}$$

$$\rho^{(1)}(r,t) = 2\operatorname{Re}\left(tr\left[\psi^{(1)}(r,t)\psi^{(0)\dagger}(r,t)\right]\right), \tag{s7}$$

$$\rho^{(2)}(r,t) = \rho^{(2)}_{02}(r,t) + \rho^{(2)}_{11}(r,t) \tag{s8}$$

with

$$\rho^{(2)}_{02}(r,t) = 2\operatorname{Re}\left(tr\left[\psi^{(0)}(r,t)\psi^{(2)\dagger}(r,t)\right]\right) \tag{s9}$$

and

$$\rho^{(2)}_{11}(r,t) = tr\left[\psi^{(1)}(r,t)\psi^{(1)\dagger}(r,t)\right], \tag{s10}$$

and

$$\rho^{(3)}(r,t) = 2\operatorname{Re}\left(tr\left[\psi^{(3)}(r,t)\psi^{(0)\dagger}(r,t)\right] + tr\left[\psi^{(2)}(r,t)\psi^{(1)\dagger}(r,t)\right]\right). \tag{s11}$$

As explained in the main text, the zeroth order term $\rho^{(0)}(r,t)$ always shows supercollimation with the initial charge density. The first order term $\rho^{(1)}(r,t)$ and the third order term $\rho^{(3)}(r,t)$ is zero since the even order terms such as $\psi^{(0)}(r,t)$ and $\psi^{(2)}(r,t)$ are orthogonal to the odd order terms such as $\psi^{(1)}(r,t)$ and $\psi^{(3)}(r,t)$ in pseudo-spin space. Among the two second order terms, the $\rho^{(2)}_{02}(r,t)$ term doesn't disrupt supercollimation because the amplitude of $\psi^{(0)}(r,t)$ is $\sqrt{\rho^{(0)}(r,t)}$, which constrains $\rho^{(2)}_{02}(r,t)$ to have the same extent and motion as $\rho^{(0)}(r,t)$ in coordinate space.

Only $\rho^{(2)}_{11}(r,t)$ may show electron dynamics deviating from electron beam supercollimation. With an initial wave packet of $\begin{pmatrix}1\\0\end{pmatrix}' \psi_0(r)\exp(i\mathbf{k}_0 \cdot r)$ where $\mathbf{k}_0$ is the initial center of mass wavevector, $\rho^{(2)}_{11}(r,t)$ is

$$\rho^{(2)}_{11}(r,t) = \int_{x-v_0 t}^{x+v_0 t} dx_1 \int_{x-v_0 t}^{x+v_0 t} dx_2 [(k_{0y} - i\partial_y)\psi_0(x_1, y)][(k_{0y} - i\partial_y)\psi_0(x_2, y)]^* \quad (s12)$$
$$\times \exp(ik_x(x_1 - x_2))\beta(x_1, x_2),$$

with

$$\beta(x_1, x_2) = \alpha(x_1, x_2)\, \alpha((x_2 + x + v_0 t)/2, (x_1 + x + v_0 t)/2)^2. \quad (s13)$$

To obtain $\rho^{(2)}_{11}(r,t)$ on average, we estimated $\overline{\beta(x_1, x_2)}$. If $|x + v_0 t| > r_0 + l_s$, the two factors in Eq. (s13) are not correlated so that

$$\overline{\beta(x_1, x_2)} \approx \overline{\alpha(x_1, x_2)}\; \overline{\alpha((x_2 + x + v_0 t)/2, (x_1 + x + v_0 t)/2)^2} \quad (s14)$$
$$\approx \overline{\alpha(x_1, x_2)}\; \overline{\alpha(x_2/2, x_1/2)^2}.$$

If we assume that $\overline{\alpha(x,x')}$ decays with a full-width-half-maximum of $l_s$ and $l_s$ is inversely proportional to the disorder fluctuation as is the case of spatially-correlated Gaussian disorder, then

$$\overline{\alpha(x_1, x_2)}\overline{\alpha(x_2/2, x_1/2)^2} \approx \overline{\alpha(x_1, x_2)}^2. \quad (s15)$$

And

$$\overline{\rho_{11}^{(2)}(\boldsymbol{r},t)} \approx \int_{x-v_0t}^{x+v_0t} dx_1 \int_{x-v_0t}^{x+v_0t} dx_2 [(k_{0y}-i\partial_y)\psi_0(x_1,y)][(k_{0y}-i\partial_y)\psi_0(x_2,y)]^* \\ \times \exp(ik_x(x_1-x_2))\overline{\alpha(x_1,x_2)}^2. \quad (s16)$$

Eq. (s16) can be simplified for $r_0 > l_s$. In that case, $\overline{\alpha(x_1,x_2)}^2$ may be approximated by $l_s\delta(x_1-x_2)$ since $\psi_0(\boldsymbol{r})$ is a smooth function. Then

$$\overline{\rho_{11}^{(2)}(\boldsymbol{r},t)} \approx l_s \int_{x-v_0t}^{x+v_0t} dx \, |(k_{0y}-i\partial_y)\psi_0(x_1,y)|^2. \quad (s17)$$

This corresponds a strip of density of width determined by the initial wavepacket but extended from $v_0t$ to $-v_0t$ in the $x$ direction. For example, for an initial Gaussian wave packet given by Eq. (6) in the main text,

$$\overline{\rho_{11}^{(2)}(\boldsymbol{r},t)} \approx \frac{l_s}{2\sqrt{2\pi}r_0}\left(k_{0y}^2 + \frac{y^2}{4r_0^4}\right)e^{-y^2/2r_0^2}\left(-\text{Erf}\left(\frac{x-v_0t}{\sqrt{2}r_0}\right) + \text{Erf}\left(\frac{x+v_0t}{\sqrt{2}r_0}\right)\right). \quad (s18)$$

Also, we shall assume that Eq. (s18) approximates $\overline{\rho_{11}^{(2)}(\boldsymbol{r},t)}$ even if $|x+v_0t| \le r_0 + l_s$ and $r_0 > l_s$. This is justified by the numerical wavepacket simulation with spatially-correlated Gaussian disorder potential described in the main text.

3. Renormalization of dispersion relation due to 1D disorder potential.

The ensemble average dispersion relation of the Dirac fermions in graphene in the presence of a 1D random potential is strongly renormalized anisotropically for $|k_y| < 1/l_s$. The contribution of the perturbing Hamiltonian, $H_1' = U^\dagger H_1 U = -v_0 p_y \sigma_y$, to the electronic states can be evaluated by using a perturbation expansion of the retarded Greens function to first order in $H_1'$,

$$G'(\mathbf{k},\mathbf{k}',\omega) \approx G'_0(k_x,k'_x,\omega)\delta_{k_y k'_y} + \sum_{k''_x} G'_0(k_x,k''_x,\omega)H'_1(k_y)G'_0(k''_x,k'_x,\omega)\delta_{k_y k'_y} \quad \text{(s19)}$$

where $H'_1(k_y) = \langle \mathbf{k}|H'_1|\mathbf{k}\rangle$. For a random potential $V(x)$, an ensemble average of Eq. (s19) should be performed. The magnitude of the off-diagonal components of $\overline{G'_0(k_x)}H_1(k_y)$ in pseudo-spin indices may be viewed as a measure of the perturbation strength on average since $\overline{G'_0}$ is a diagonal matrix with respect to pseudo-spin indices as well as with respect to wavevectors (since translation symmetry of the pristine system is restored by ensemble average [2]) and the diagonal components of $H'_1(k_y)$ in the pseudo-spin indices are zero. The maximum value of $|\overline{G'_0}(k_x,\omega)|$ is $\sim \tau/\hbar I'$ if we assume that the spectral function of an electronic state with $k_x$ and $s$ follows a Lorentzian distribution centered at $\omega = sv_0 k_x$ with a full width at half maximum of $\sim 2\hbar/\tau$. Since $H'_1(k_y) = -\hbar v_0 k_y \sigma_y$, the first and higher order contribution from $H'_1$ can be neglected if $|\tau v_0 k_y| = |k_y l_s| < 1$, resulting in a flat dispersion given by $H'_0$.

4. Supercollimation enhancement in graphene superlattices by additional 1D disorder potential.

By disordering the external periodic potential of a graphene superlattice that does not give supercollimation, we can yield supercollimation. We demonstrate this effect numerically using a width-height-center disorder to a particular type of periodic potential with a height of $U_0$ and a period of $l$ shown in Fig. S1(a). For each potential unit, we can disorder its center, width, and potential height by using mutually independent Gaussian random variables with a standard deviation of $\Delta_D$ centered at the original value. We calculated wavepacket propagation of the Dirac fermions under this disordered periodic potential from 60 ensemble average with $U_0 = 2\pi\hbar v_0 / l$ (which is half of the potential magnitude needed for supercollimation in SGS) and an $\Delta_D$ that is a 100 % of the original variables (potential height, center and width). The initial wave packet is in a Gaussian shape with initial center of mass wavevector $k_0 = \pi / 5l_c$ and a half width of $r_0 = 5l_c$. Figure S1(c) and S1(d) show the electron density distributions at two different incident angles $\theta$ of 0° and 45° from the x axis. Without disorder, the electrons propagate nearly along the incident center of mass wavevector direction. Although the electron beam is slightly collimated along the $x$ direction (owing to the superlattice potential) compared to the pristine system shown in Fig. 1 in the main paper, it still spreads along the y direction. If we add the above disorder to this periodic potential, the Gaussian wave packet maintains its shape and propagates along the direction of the external potential modulation regardless of the incident angle.

We may also demonstrate this effect by evaluating $\overline{\rho_{11}^{(2)}(\boldsymbol{r},t)}$ in Sec. 2 above since the other terms to $\rho(\boldsymbol{r},t)$ up through third order always show supercollimation as discussed above. For a system under disorder potential in addition to the periodic potential, if $|x + v_0 t| > r_0 + l_s$,

$$\overline{\rho_{11}^{(2)}(\boldsymbol{r},t)} \approx \int_{x-v_0 t}^{x+v_0 t} dx_1 \int_{x-v_0 t}^{x+v_0 t} dx_2 [(k_{0y} - i\partial_y)\psi_0(x_1, y)][(k_{0y} - i\partial_y)\psi_0(x_2, y)]^* \\ \times \exp(ik_x(x_1 - x_2))\overline{\alpha_D(x_1, x_2)\alpha_D(x_2/2, x_1/2)^2}\beta_P(x_1, x_2). \quad \text{(s20)}$$

Here $\beta_P(x_1, x_2)$ is $\beta(x_1, x_2)$ of the periodic potential and $\alpha_D(x_1, x_2)$ is $\alpha(x_1, x_2)$ of the disorder potential. If we assume that $\overline{\alpha_D(x, x')}$ decays with a full-width-half-maximum of $l_s$ and $l_s$ is inversely proportional to the disorder fluctuation as is in the case of spatially-correlated Gaussian disorder, then

$$\overline{\alpha_D(x_1, x_2)\alpha_D(x_2/2, x_1/2)^2} \approx \overline{\alpha_D(x_1, x_2)}^2. \quad \text{(s21)}$$

And we have

$$\overline{\rho_{11}^{(2)}(\boldsymbol{r},t)} \approx \int_{x-v_0 t}^{x+v_0 t} dx_1 \int_{x-v_0 t}^{x+v_0 t} dx_2 [(k_{0y} - i\partial_y)\psi_0(x_1, y)][(k_{0y} - i\partial_y)\psi_0(x_2, y)]^* \\ \times \exp(ik_x(x_1 - x_2))\overline{\alpha_D(x_1, x_2)}^2 \beta_P(x_1, x_2). \qquad (s22)$$

For $r_0 > l_s$, $\overline{\alpha_D(x_1, x_2)}^2$ may be approximated by $l_s \delta(x_1 - x_2)$ since $\psi_0(\boldsymbol{r})$ is a smooth function. Then,

$$\overline{\rho_{11}^{(2)}(\boldsymbol{r},t)} \approx l_s \int_{x-v_0 t}^{x+v_0 t} dx \, |(k_{0y} - i\partial_y)\psi_0(x_1, y)|^2. \qquad (s23)$$

Here, we assume that Eq. (s23) approximates $\overline{\rho_{11}^{(2)}(\boldsymbol{r},t)}$ even if $|x + v_0 t| \leq r_0 + l_s$ and $r_0 > l_s$. This is justified by the numerical wavepacket simulation with center-width-height disorder potential described in the main text. Eq. (s23) is the same form as Eq. (s17), indicating that $\overline{\rho_{11}^{(2)}(\boldsymbol{r},t)}$ decreases as $l_s / r_0$ decreases as shown in Eq. (s18).

5. Spatially correlated Gaussian disorder potential generation.

We generated the spatially correlated Gaussian disorder in the simulation by using a Cholesky decomposition of the two-point correlation matrix [3]. First, we generated a vector, $V_i = V(x_i)$, consisting of spatially-uncorrelated Gaussian-random variables having zero mean and a variance of 1. $V_i$ can be characterized by $\overline{\underset{\sim}{V}\underset{\sim}{V}^T} = \underset{\sim}{I}$, where the tilde represents a matrix or vector and $\underset{\sim}{I}$ represents an identity matrix. A random vector $\underset{\sim}{W}$ with a desired spatial correlation, $\overline{\underset{\sim}{W}\underset{\sim}{W}^T} = \underset{\sim}{C}$ where $C_{ij} = \Delta^2 e^{-|x_i - x_j|/l_c}$, can be obtained using by a Cholesky decomposition of $\underset{\sim}{C} = \underset{\sim}{L}\underset{\sim}{L}^T$ that is symmetric and positive definite by definition. If we construct a random vector $\underset{\sim}{W} = \underset{\sim}{L}\underset{\sim}{V}$ using the above generated $\underset{\sim}{V}$ and $\underset{\sim}{L}$ then

$$\overline{\underset{\sim}{W}\underset{\sim}{W}^T} = \overline{\underset{\sim}{L}\underset{\sim}{V}\underset{\sim}{V}^T\underset{\sim}{L}^T} = \underset{\sim}{L}\overline{\underset{\sim}{V}\underset{\sim}{V}^T}\underset{\sim}{L}^T = \underset{\sim}{L}\underset{\sim}{L}^T = \underset{\sim}{C} \qquad (s24)$$

Hence, the random vector $\underset{\sim}{W} = \underset{\sim}{L}\underset{\sim}{V}$ would have the desired two point correlation matrix of $\underset{\sim}{C}$.

6. Conductance calculation.

The conductance $G$ is calculated using Eq. (12) in the main text. Suppose that we expand $G^{R-A}$ and conductance $G$ to first order in $H_1'$

$$G(L,\theta) \approx G^{(0)}(L,\theta) + G^{(1)}(L,\theta) + \cdots \quad (s25)$$

$$G^{R-A}(r,t) \approx G^{R-A,(0)}(r,t) + G^{R-A,(1)}(r,t) + \cdots \quad (s26)$$

where conductance $G^{(i)}$ and $G^{R-A,(i)}$ represent the ith order term in $H_1'$. Substituting these expansions to Eq. (12) in the main text and arranging them according to powers in $H_1'$, we have

$$G^{(0)}(L,\theta) = \frac{\pi\hbar}{(2\pi i)^2}$$
$$\int dr dr' \, tr\{J_\mu(r)(G^{R-A,(0)}(r,r',E_F))J_\mu(r')(G^{R-A,(0)}(r',r,E_F))\}\delta(\mu)\delta(\mu'-L), \quad (s27)$$

and

$$G^{(1)}(L,\theta) = \frac{\pi\hbar}{(2\pi i)^2}$$
$$\int dr dr' \, \Big[tr\{J_\mu(r)(G^{R-A,(0)}(r,r',E_F))J_\mu(r')(G^{R-A,(1)}(r',r,E_F))\}$$
$$+tr\{J_\mu(r)(G^{R-A,(1)}(r,r',E_F))J_\mu(r')(G^{R-A,(0)}(r',r,E_F))\}\Big]\delta(\mu)\delta(\mu'-L). \quad (s28)$$

The zeroth order term in momentum space is, if we assume that $\overline{\alpha(|x_1-x_2|)}$ decays exponentially,

$$\overline{G^{(0)}(L,\theta)} \approx \frac{2e^2}{h}\sum_{k_v}\left(\cos^2\theta + \sin^2\theta\exp[-2L/l_s\cos\theta]\cos(2E_F L/(\hbar v_0 \cos\theta))\right). \quad (s29)$$

with periodic boundary condition along the $v$ direction. The first order term $G^{(1)}(L,\theta) = 0$ due to cancellation of the two terms in Eq. (s28)

7. Upper bound for disorder correlation length

In this section, we discuss an additional condition for observing supercollimation using an external disorder potential: the upper bound for disorder correlation length $l_c$.

Firstly, $l_c$ should be shorter than the phase coherence length $L_\phi$ of the system. Phase coherence length is the length scale beyond which there is no interference effect or no phase coherence [4]. It is originated from the incoherent and irreversible processes due to coupling of an electron to the environment such as interaction with phonons, other electrons, and impurities with internal degrees of freedom. For the case of pristine graphene, it is $\sim 0.5\,\mu m$ [5]. We may expect that for a sample with $l_c \gg L_\phi$, an electron would see a more or less constant potential before losing its phase-coherence. In this case, the broadening in the spectral function is originated from the distribution of the constant potential heights, not from the random phase accumulation in each member of the ensemble. In contrast, if $l_c < L_\phi$, the electron sees the potential variation before losing phase coherence and the external potential plays a role of a disorder potential. Then, broadening in the spectral function is originated from the random phase accumulation of the electron.

Secondly, there is another consideration on the upper limit in $l_c$ set by the reflection of a charge carrier from a slowly-varying potential. It is known that if an electronic state in graphene with wavevector $\boldsymbol{k}_0$ encounters a slowly-varying p-n junction such that the external potential changes by $\hbar v_0 k_0$ over a distance $d_c$, the propagating state will be partially reflected from the junction if $k_0 d_c > 1$ [6]. We have considered a potential with characteristic variation in strength by $\Delta$ over a distance of $\sim l_c$; hence, in our case, the distance $d_c$ for partial reflection corresponds roughly to $d_c \sim l_c \frac{\hbar v_0 k_0}{\Delta}$ and the condition that such reflection does not occur is

$k_0 \left( l_c \frac{\hbar v_0 k_0}{\Delta} \right) < 1$. Or, if we define $L_{PN} = \Delta / (\hbar v_0 k_0^2)$, the second condition on $l_c$ reduces to $l_c < L_{PN}$.

Combining these two effects, the upper bound on $l_c$ for observing supercollimation discussed in our paper is roughly given by the shorter of the two lengths, $L_\phi$ and $L_{PN}$.

8. Upper bound for the random phase accumulation length $l_s$

For observing the supercollimation phenomenon, the standard elastic mean free path due to usual isotropic distribution of static scatterers such as charged impurities for the case of regular graphene also provides an upper bound for the value of $l_s$ of the external 1D disorder potential.

9. Supercollimation in graphene with tight-binding model.

The discussions and calculations above are done based on the low-energy Dirac Hamiltonian given by Eq. (1) in the main text. We have also carried out simulations using a $\pi$-band tight-binding Hamiltonian with nearest-neighbor hopping integral ($t$ =2.7 eV) for graphene with an on-site Gaussian disorder potential fluctuating along the x direction shown in Fig. S2 (a) below. We used in the simulations a graphene supercell that is $1\mu m$ long along the x direction. Figure S2 (b)-(d) demonstrate supercollimation in a Gaussian wavepacket propagation simulation within this tight-binding framework. Using 45 different realizations of an onsite spatially-correlated Gaussian disorder potential $V(x)$ with a disorder magnitude $\Delta = t/8 \sim 0.3 eV$ and a correlation length $l_c = 20nm$, we numerically calculated the electron density $\overline{\rho(\mathbf{r},t)}$ using an initial Gaussian packet with center of mass wavevector energy corresponding to $E - E_D = 20meV$ and a half width of $r_0 = 80nm$. Figure S2 (c) and S2 (d) show the evolution of the electron density $\overline{\rho(\mathbf{r},t)}$ from the initial electron density shown in Fig. S2 (b). In the absence of a disorder potential, the Gaussian wavepacket propagates along the initial center of mass wavevector direction marked by the white arrow and spreads sideway. With the 1D disorder potential $V(x)$, the electron package propagates virtually un-spread along the potential fluctuation direction, which is $x$, regardless of the initial packet velocity direction.

**Figures**

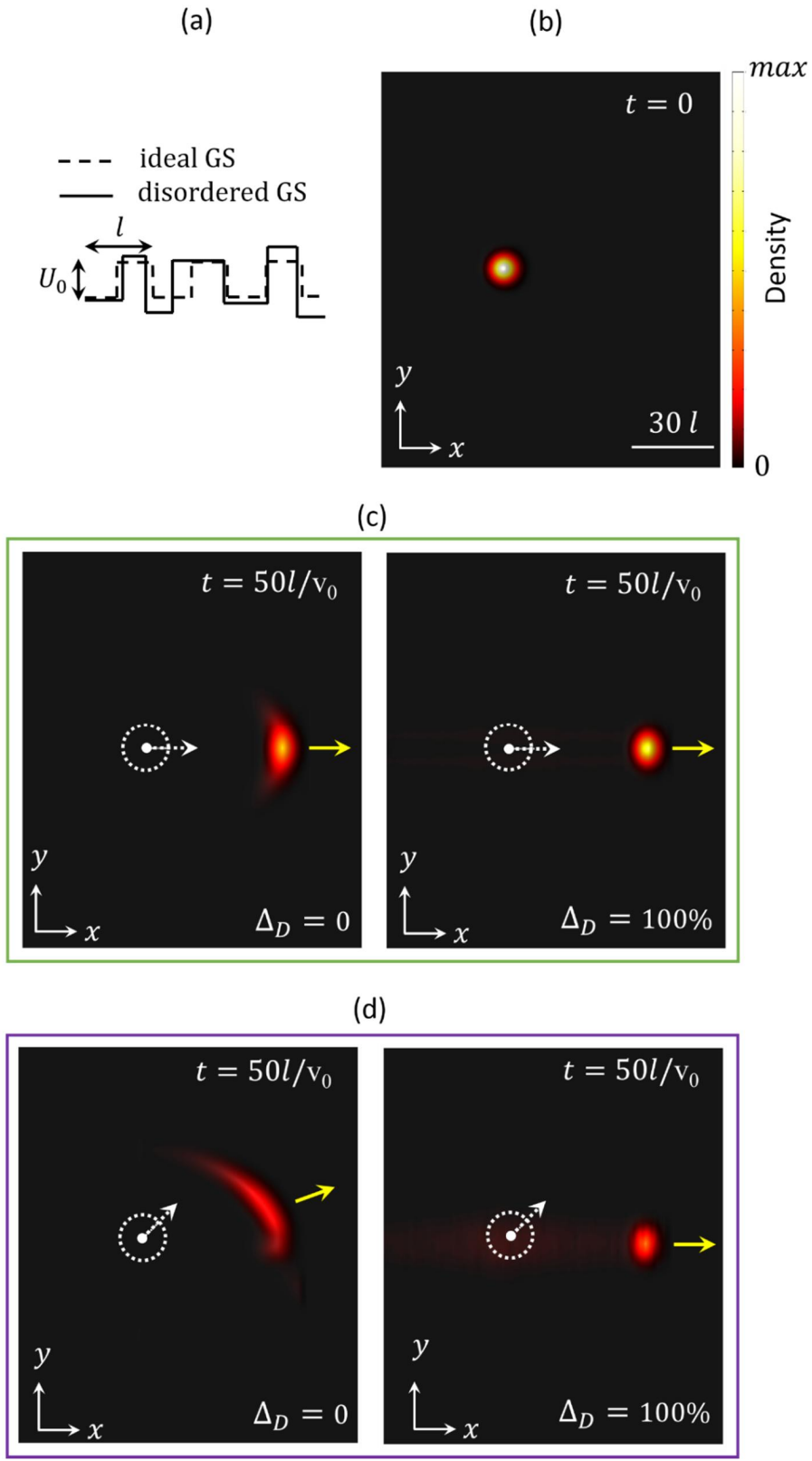

Fig. S1. (a) A realization of a periodic potential with 1D disorder. The full and dashed lines represent disordered and ideal periodic external potential, respectfully. (b) Initial wavepacket with electron density in a Gaussian shape in coordinate space with initial center of mass wavevector of $k_0 = \pi/5l$ and $r_0 = 5l$. (c)-(d) Electron density distribution in coordinate space at $t = 50l/v_0$ in graphene under periodic potential with $U_0 l = 2\pi\hbar v_0$ (left panel) and in graphene under the same periodic potential but with disordered height, center and width of $\Delta_D = 100\%$ (right panel) with initial center of mass wavevector direction (white arrow) pointing with respect to the $\hat{x}$ axis at $0°$ (c) and $45°$ (d).

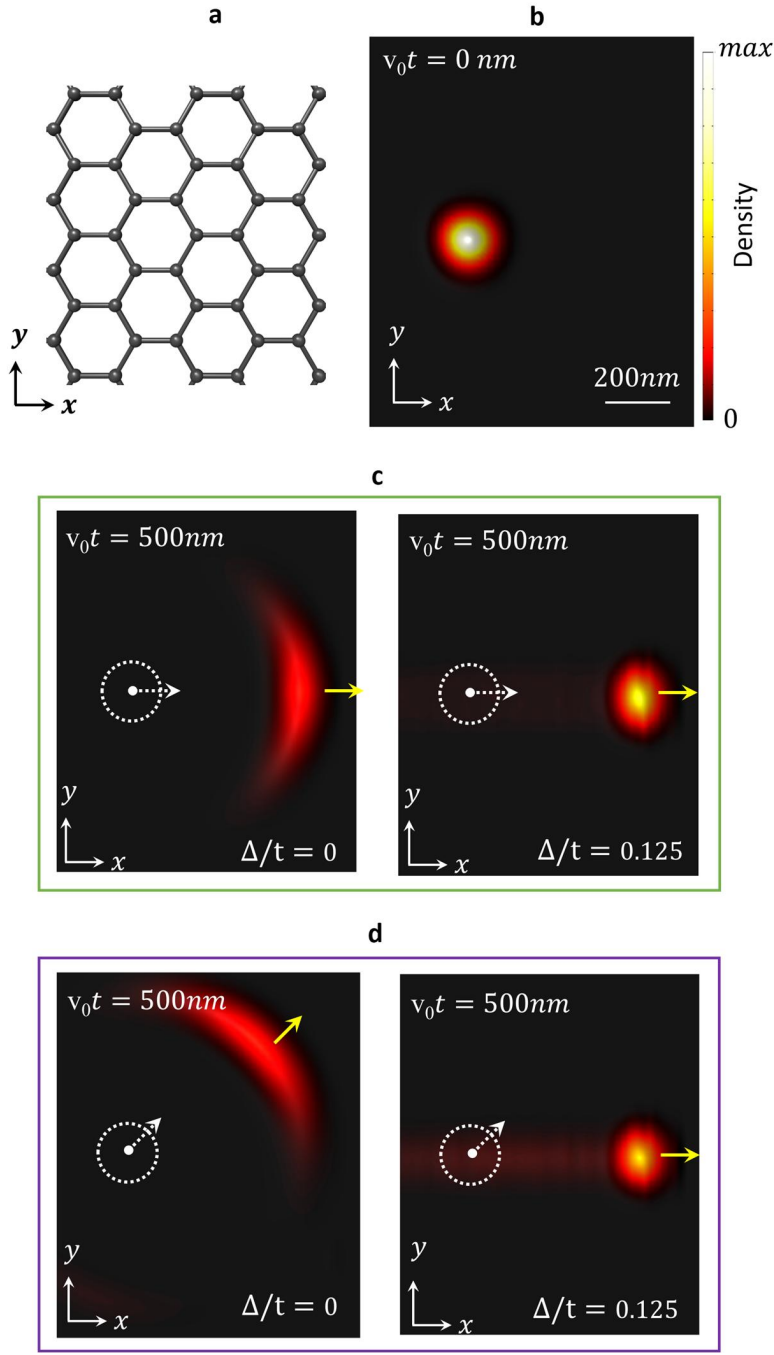

Fig. S2. (a) Atomic structure of graphene. External disorder potential is taken to be an onsite spatially-correlated Gaussian disorder potential fluctuating along the x direction. (b) Initial wavepacket with electron density in a Gaussian shape in coordinate space with center-of-mass wavevector corresponding to an energy of $E - E_D = 20 meV$ and a half width of $r_0 = 80 nm$,

where $E_D$ is the Dirac point energy. (c)-(d) Electron density distribution in coordinate space at $v_0 t = 500 nm$ in the pristine system (left panel) and in the disordered system (right panel) with initial center of mass wavevector direction (white arrow) pointing with respect to the x axis at $0°$ (c) and $45°$ (d).